\documentclass[conference]{IEEEtran}
\usepackage{fourier} %% nicer font

\usepackage{color}
\usepackage{graphicx}
\usepackage{caption}
\usepackage{multirow}
\usepackage{hyperref}
\usepackage{fancyvrb}
\usepackage{relsize}
\usepackage{listings}
\usepackage{listings,tikz,array}
\usepackage{float}

\usepackage[T1]{fontenc}  
\usepackage[scaled=0.82]{beramono}  

\usepackage{microtype} 

\sbox0{\small\ttfamily A}
\edef\mybasewidth{\the\wd0 }

\lstdefinelanguage{scala}{
  morekeywords={abstract,case,catch,class,def,%
    do,else,extends,false,final,finally,%
    for,if,implicit,import,match,mixin,%
    new,null,object,override,package,%
    private,protected,requires,return,sealed,%
    super,this,throw,trait,true,try,%
    type,val,var,while,with,yield},
  sensitive=true,
  morecomment=[l]{//},
  morecomment=[n]{/*}{*/},
  morestring=[b]",
  morestring=[b]',
  morestring=[b]"""
}

\usepackage{color}
\definecolor{dkgreen}{rgb}{0,0.6,0}
\definecolor{gray}{rgb}{0.5,0.5,0.5}
\definecolor{mauve}{rgb}{0.58,0,0.82}

\lstnewenvironment{scala}[1][]
{\lstset{language=scala,#1}}
{}
\lstnewenvironment{cpp}[1][]
{\lstset{language=C++,#1}}
{}
\lstnewenvironment{bash}[1][]
{\lstset{language=bash,#1}}
{}
\lstnewenvironment{verilog}[1][]
{\lstset{language=verilog,#1}}
{}

\DefineVerbatimEnvironment%
    {quoze}%
    {Verbatim}%
    {xleftmargin=2em, samepage=true, fontsize=\relsize{-2}, frame=single}

    \DefineVerbatimEnvironment%
        {quoze2}%
        {Verbatim}%
        {xleftmargin=5em, samepage=true, fontsize=\relsize{-3}, frame=single}

\ifCLASSINFOpdf
\else
\fi

\usepackage[caption=false,font=footnotesize]{subfig}

\newbox\verbbox

\begin{document}

\title{Research Note: An Open Source Bluespec Compiler}

\author{David~J.~Greaves
 --- March 2013 (2$^{nd}$ ed.~1Q19)}
% The date is that of the main body of work, where the cth-tiny cpu was completed on 20th Jan 2013.
\date{March 2013} 

\maketitle

\begin{abstract}

In this Research Note we report on an open-source compiler for the Bluespec hardware
description language.

\end{abstract}

% \fi
%

\IEEEpeerreviewmaketitle

% \noindent {\bf Keywords:} Dark Silicon, Debug and Trace, on-chip debug support, Energy efficient computing, spEEDO.

% \noindent {\bf THIS IS AN EARLY DRAFT OF A PAPER WHERE THE WORK IS NEARLY COMPLETE.}

% {\bf THIS IS A PRE-PRINT OF A PAPER TO BE PRESENTED AT ?? .}

\section{Introduction}

Warning: This note is based on limited experience with the Bluespec language and may embody
a few misapprehensions.

Bluespec \cite{bluespec:04} is a programming language for generating hardware circuits.
The Bluespec language was created at MIT and is now promoted by Bluespec Inc.
The compiler from that company is only available under license.

Although there is no accepted taxonomy of high versus low-level
languages for hardware design, we can roughly relate a gate-level netlist
to machine code, RTL to assembly language, hardware construction
languages such as Chisel\cite{Chisel-Bachrach-Dac2012} and Lava\cite{Lava1998} as low-level languages and anything
that makes automatic assignment of work to clock cycles as high-level
languages.  Accordingly, Bluespec can be classed as a high-level
language. However, it arguably sits at a lower level
than traditional HLS (high-level synthesis) since Bluespec
does not make heuristic-guided searches for optimal binding of operations
to functional units (FUs such as ALUs and RAMs) or multi-cycle static schedules.

Programs in a `Hardware Construction Language', such as Chisel,
essentially `print out' an RTL or structural design.  This process is
called {\tt structural elaboration}. HardCaml, Clash and Lava \cite{Lava1998} are
further examples.  The {\tt generate} statements of Verilog and VHDL
form the hardware construction languages of those RTLs.  Bluespec
embodies a sophisticated hardware construction language based on
functional programming combinators.  The structural elaboration may
contain loops and other control flow constructs, but the elaboration
is performed entirely at compile time. Hence none of the conditional
statements processed in the hardware construction language depends on
any run-time data. There is no data-dependent control flow in the
elaboration language.

Bluespec is based around the concept of modules and rules.
A module contains zero or more rules. A module also instantiates zero or more lower modules.
Modules instantiated at the lowest levels are primitives, such as FIFOs, registers and RAMs.
Bluespec starts structural elaboration at a top-level module.
The module hierarchy is nominally flattened
during the structural elaboration process.  Once  elaboration is
complete, we have essentially a flat collection of interconnected Bluespec
rules and primitives.

The standard compilation semantics for Bluespec enforce a particular
mapping between rule firing and hardware clock cycles, such as a
register only being updated by exactly one firing of at most one rule
in any clock cycle.

Where the design hierarchy is partitioned into separate compilation
units, which can be done with compiler directives or annotations embedded
in the source code, there is a variation in semantic of interaction between modules within the
compilation unit and those in different units (methods are not re-entrant when invoked
from a separate compilation unit).

\setbox\verbbox=\vbox{\hsize=3in
\tiny   \begin{Verbatim}
module mkTb1 (Empty);

   Reg#(int) x <- mkReg (23);

   rule countup (x < 35);
      int y = x + 1;      // This is short for  int y = x._read() + 1;
      x <= x + 1;         // This is short for  x._write(x._read() + 1);
      $display ("x = %0d, y = %0d", x, y);
   endrule

   rule done (x >= 30);
      $finish (0);
   endrule

endmodule: mkTb1
  \end{Verbatim}
}

\begin{figure}\center
  \begin{tabular}{|c|}
      \hline
     \box\verbbox \\
      \hline
      \end{tabular}
\caption{A short, flat Bluespec program with two rules sharing one register. \label{fig:listingshort2}}
\end{figure}

Figure~\ref{fig:listingshort2} presents a small example with two rules: one called \verb+countup+ increments, the other, called \verb+done+,
exits the simulation.  
% This example looks very much like RTL: provides an easy entry for hardware engineers.
A potential problem with this example is that `clean' atomic rules are acting on a common, shared variable (the register) and the resulting
behaviour might not be `clean' in that a predictable result requires a schedule with predictable interleaving.
Since this example has low complexity, the designer
can be readily confident that the \verb+done+ rule is schedulled sufficiently often for its body to be executed the moment \verb+x+ gets
to 30, but in a more complicated design, other rules might have higher priority and outcomes will depend on fairness
rules.  An attribute, \verb+fire_when_enabled+, could have been added to cause an error if the scheduller cannot guarantee
the expected behaviour.

In general, RAMs and registers shared by freely-schedullable rules will suffer from RaW (read-after-write) hazards and the like.
Detailed behaviour will depend on the schedulling chosen. 
% Cleanliness is preserved when rules communicate with FIFOs, like CSP process networks.

\begin{figure}[htb]
%  \centerline{\includegraphics[width=8 cm]{images/toy-bsv-flow.eps}}
\centerline{\includegraphics[width=8 cm]{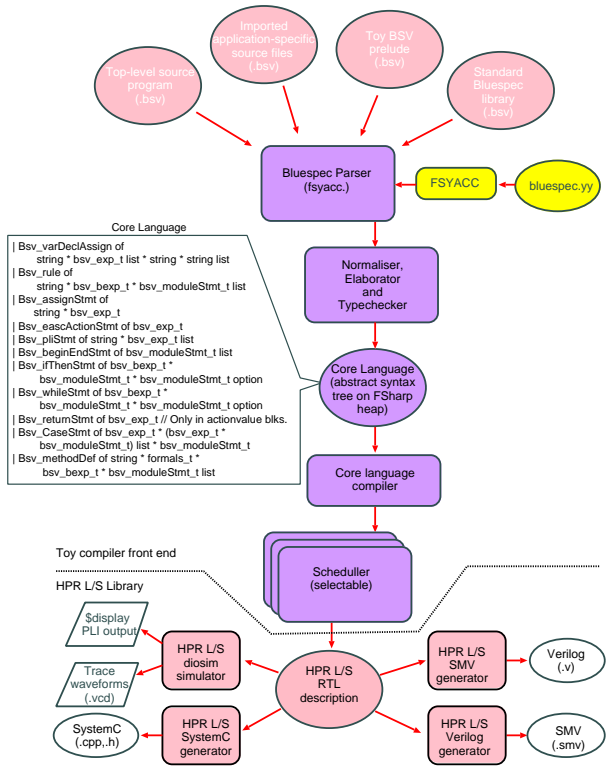}}  
\caption{Overall structure of the Open Source Compiler.) \label{fig:bsvc-hprls}}
\end{figure}

{\em Why write an open-source compiler?}  1.~This exercise was largely conducted to understand
more about the language. But the result is useful to those who want to try out the language
for simple experiments or small-scale teaching without licensing the commercial version.
2.~A second-source for a proprietary resource can promote its adoption by those who do not
want to enter into code escrow.  3.~We wanted a testbed to experiment with alternative approaches
to rule schedulling and meeting hard real-time performance goals.

Our first implementation was just a proof of concept, without a parser or structural elaborator.
Instead we manually entered the abstract syntax tree of the elaborated modules
as an FSharp data structure.  This worked well enough, so we proceeded to our second
implementation. This added a front-end parser and elaborator. Implementing these
and getting them to work on public domain Bluespec examples
was significantly more work than the core language compiler itself.

\section{Core Language Abstract Syntax}

After the elaboration phase, we have the executable Bluespec program expressed using the following abstract syntax.

{\small\begin{verbatim}
type bsv_provisos_t =
| Proviso_alwaysReady
| Proviso_alwaysEnabled


// Method prototype protocol has one of these three forms:
type bsv_protoprotocol_t =
| BsvProtoValue of bsv_type_name_t
| BsvProtoActionValue of bsv_type_name_t
| BsvProtoAction

and bsv_type_name_t =
| Bsv_typen of string    // A named type
| Bsv_typen_uint_ns      // Native/primitive run-time type
| Bsv_typen_uint1 of int // Native/primitive run-time type
| Bsv_typen_action

// Method prototype: has name, provisio/pragmas, return type
// and method formals (pairs of formal and actuals so-far
// bound).
and bsv_methodProto_t = Bsv_methodProto of string *
  bsv_provisos_t list * bsv_protoprotocol_t *
  (bsv_type_name_t * string) list

and bsv_sigma_protocol_t =
| BsvValue
| BsvAction of hexp_t option * (hbexp_t * hexp_t)

and bsv_sigma_t = Bsv_current of string *
       bsv_sigma_protocol_t * (bsv_type_name_t *
       hexp_t option) * (hexp_t * hexp_t) list

and actuals_t = (string * (bsv_type_name_t *
       bsv_exp_t)) list

and formals_t = (bsv_exp_t * string) list

and bsv_moduleStmt_t = // Or actionValueStmt ?
| Bsv_varDeclAssign  of string * bsv_exp_t list * string *
       string list
| Bsv_rule           of string * bsv_bexp_t *
       bsv_moduleStmt_t list
| Bsv_assignStmt     of string * bsv_exp_t
| Bsv_eascActionStmt of bsv_exp_t
| Bsv_pliStmt   of string * bsv_exp_t list //$display etc
| Bsv_beginEndStmt   of bsv_moduleStmt_t list
| Bsv_ifThenStmt     of bsv_bexp_t * bsv_moduleStmt_t *
       bsv_moduleStmt_t option
| Bsv_whileStmt     of bsv_bexp_t * bsv_moduleStmt_t *
       bsv_moduleStmt_t option
| Bsv_returnStmt    of bsv_exp_t // actionvalue blocks only.
| Bsv_CaseStmt      of bsv_exp_t * (bsv_exp_t *
       bsv_moduleStmt_t) list *  bsv_moduleStmt_t option
| Bsv_methodDef     of string * formals_t * bsv_bexp_t *
       bsv_moduleStmt_t list
| Bsv_primBuffer    of bsv_exp_t * bsv_exp_t
\end{verbatim}}

The expression syntax is a relatively straightforward grammar with the most interesting aspect being the {\tt B\_apply} construct:
{\small \begin{verbatim}
and bsv_moduleParams_t = (bsv_type_name_t * string) list *
   (bsv_type_name_t * string) list * bsv_type_name_t

// The Bluespec interface definition: Generic
//  parameters and list of methods:
and bsv_if_t = Bsv_if of (bsv_type_name_t * string) list
        *  bsv_methodProto_t list

// Boolean expressions
and bsv_bexp_t =
| B_true
| B_false
| B_firing of string // Backdoor access
        to the composite guard for any rule.
| B_not  of bsv_bexp_t
| B_and  of bsv_bexp_t list
| B_or   of bsv_bexp_t list
| B_bexp of hbexp_t // Pre-compiled forms
| B_bdiop of x_bdiop_t * bsv_exp_t list // Equality,
       less than and orreduce where list is a singleton.
| B_orred of bsv_exp_t

// Integer expressions
and bsv_exp_t =
| B_num of int list
| B_blift of bsv_bexp_t
| B_query of bsv_bexp_t * bsv_exp_t * bsv_exp_t
| B_var of string
| B_string of string
| B_hexp of hexp_t
| B_fsmStmt of bsv_exprFsmStmt_t
| B_ifcRef of string // Interface reference
| B_diadic of x_diop_t * bsv_exp_t * bsv_exp_t
| B_apply of string list * bsv_exp_t list
\end{verbatim}}

The apply construct either maps to a built-in function, such as \verb+sizeof+ or
invokes a method on a primitive component. The method must return its result within the same clock cycle (we return to this later).

\subsection{FSM sub-language}
Finally there is the finite-state machine sub-language that
defines an FSM with a state variable, transition rules and actions.
The FSM sub-language is syntactic sugar and is
replaced with standard Bluespec rules in a front-end stage.
See function {\tt norm\_subl} which is approximately 50 line of FSharp starting at line 9155 of {\tt bsvc.fs}.
It would potentially be useful to preserve and exploit the time-domain disjointedness of the FSM states when schedulling, but since our implementation is
built on the HPR L/S library \cite{HPRLS03}, this information is intrinsically captured and readily returned from
the {\tt meox.enumf} structures that symbolically represent the state enumeration.
So there is nothing to be gained from preserving the input form.

% Also, imperative expression using a conceptual thread is also much loved by programmers, so Bluespec has a behavioural sub-language compiler built in that generates state machines.

{\small \begin{verbatim}
and bsv_exprFsmStmt_t =
| Bsv_seqFsmStmt    of bsv_exprFsmStmt_t list
| Bsv_parFsmStmt    of bsv_exprFsmStmt_t list
| Bsv_ifFsmStmt     of bsv_bexp_t * bsv_exprFsmStmt_t *
        bsv_exprFsmStmt_t option
| Bsv_whileFsmStmt  of bsv_bexp_t * bsv_exprFsmStmt_t
| Bsv_repeatFsmStmt of bsv_exprFsmStmt_t
| Bsv_eascFsmStmt   of bsv_exp_t
| Bsv_breakFsmStmt
| Bsv_continueFsmStmt
\end{verbatim}

A Bluespec module definition exposes an interface called the implemented interface.
An interface is a collection of callable methods.
The top-most module in a runnable design must implement the {\tt Empty} interface that has no methods.
Where incremental compilation is used, a sub-component can be compiled that does not have an Empty interface.   As well
as the implemented interface, there can be zero or more further interfaces referenced in the module's signature.  These must be
connected up to appropriate instances of further modules by the instantiating module.  Where a module instantiates child modules, it
can use the interfaces these children expose.  From the point of view of rules in the current module,
there is no difference between using the methods of the referenced interfaces and using the methods of the instantiated instances.
There are also constructs to pass-up the interface of a child component as a component of the implemented interface.

\setbox\verbbox=\vbox{\hsize=3in
\tiny   \begin{Verbatim}
module mkTb2 (Empty);

   Reg#(int) x    <- mkReg ('h10);
   Pipe_ifc  pipe <- mkPipe;

   rule fill;
      pipe.send (x);
      x <= x + 'h10;   // This is short for  x.write(x.read() + 'h10);
   endrule

   rule drain;
      let y = pipe.receive();
      $display ("    y = %0h", y);
      if (y > 'h80) $finish(0);
   endrule
endmodule
\end{Verbatim}
}

\begin{figure}[htb]
%\centerline{\includegraphics[width=8 cm]{images/bluespecpipe.eps}}
\centerline{\includegraphics[width=8 cm]{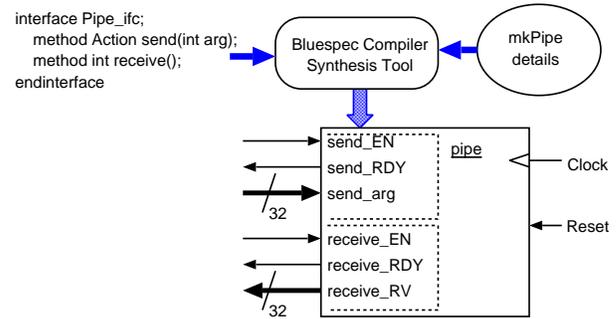}}  
\caption{Synthesis of the `pipe' Bluespec component, illustrating handshake nets. (The internal workings of the component
  are not relevant to our discussion and not presented in this report. We report only on the interfacing details.) \label{fig:bluespecpipe}}
\end{figure}

\begin{figure}\center
  \begin{tabular}{|c|}
      \hline
     \box\verbbox \\
      \hline
      \end{tabular}
\caption{Shown creation of an instance of the `pipe' component with two rules operating on it. \label{fig:listingpipe}}
\end{figure}

Nominally, every method of an interface has a direct hardware equivalent
at the net level where there are busses for each argument and for the
result, if any. At the net level there is also a bi-directional
handshake consisting of ready and enable signals. A transaction happens
at an interface on any clock cycle where both handshake nets hold.
There is no support for pipelining or processing delay within the
module that provides the interface.  When delay is inevitable, such as
in reading a synchronous RAM with latency of one or more clock cycles,
or invoking a floating-point {\sc add} that is perhaps fully-pipelined
(unity initiation interval) but which has latency of five, two methods
must be used: one that presents the input data (void result) and one
that collects the result (unit argument).  These are commonly called
{\tt put()} and {\tt get()} giving the Put/Get paradigm.

The example in Figures~\ref{fig:bluespecpipe} and~\ref{fig:listingpipe} shows a FIFO-like component called `pipe' that is acted on by two
rules.  This is immune from schedulling artefacts/hazards.  The
example interface is for a pipeline object that could have arbitrary
delay.  The sending process is blocked by implied handshaking wires
(hence far less typing than Verilog) and in the future would allow the
programmer or the compiler to re-time the implementation of the pipe
component.

The net-level representation of every method is nominal in that, for
methods both defined and used in the current compilation unit, the
nets do not need to be generated and the method is instead elaborated
as many times as it called, thereby becoming re-entrant. We call these \textit{ephemeral} methods.
Leaf components, such as RAMs and registers behave like separate
compilation units: their methods are \textit{non-ephemeral} since they
normally\footnote{As an extension, the open-source compiler enables re-entrant register use when the
  {\tt -bsv-enable-multiple-writes=enable} recipe flag is provided.} map to real hardware components.

% Bluespec RTL was intended to be declarative, both in the elaboration language and with the guarded atomic actions for actual register transfers.
% Its advanced generative elaborator is a functional language and a joy to use for advanced/functional programmers.
% So it is/was much nicer to use than pure RTL.
% It has a scheduler (cf DBMS query planner) and a behavioural-sub language for when imperative is best.

A method is invoked using the apply statement.  A simple module with empty interface containing one primitive and
one rule with two method applications looks as follows in source an abstract syntax form:
{\small \begin{verbatim}
(* module test1();
      Reg#(#UInt(15)) ctr1 <- mk_register(0)
      rule test_ast1 (True) ;
          ctr1.write(ctr1._read()+1)
      endrule
  endmodule
*)
Bsv_moduleDef("test1",
    ([], [], Bsv_typen "Empty"),
      [
        Bsv_varDeclAssign("ctr1", [B_num [15];
             B_num[1]], "mkReg", []);
        Bsv_rule("test_ast1", B_true,
          [ B_apply(["ctr1"; "_rite"], B_diadic(V_plus,
            B_apply(["ctr1"; "_read"],
            B_num[1]))])
      ],
    [])
\end{verbatim}

In this example, a primitive register is created (15-bits wide, reset
to 0). Such a register has two methods: read and write.  Since the
value written is one greater than the value read, the rule increments
the register.  Bluespec standard semantics specifies that any rule can fire at most once
per clock cycle, so the effect is that the counter can increment at
most once per clock cycle.  In fact it will increment every clock
cycle since there is nothing to stop the rule firing every clock
cycle.  In general, rules will fire if they can, but they are
prevented from firing by various interlocks and schedulling
constraints.

For a register, the method calls {\tt \_read} and {\tt \_write} are generally
inferred from syntactic sugar implemented in the Bluespec parser, but for
the sake of clarity we made them explicit in our first example.
The concise form is written  `{\tt ctr1 <= ctr1 + 1}'.

One firing constraint is the so-called `explicit guard' which is boolean expression accompanying the rule.  In this example the expression '{\tt True}' was given
as the explicate guard, and this always holds and can be left out.
A condition such as '{\tt ctr1<10}' could have been given.  This would stop the counter once it had
reached ten.  This would be expanded in the front-end parser to `{\tt B\_didic(V\_dltd, B\_apply(["ctr1"; "\_read"], []), B\_num[10])}' and hence would involve a second
method call on the counter.  The Bluespec compiler actually elaborates all of this overhead away, resulting in the RTL `{\tt if (ctr1<=10) ctr1 <= ctr1+1;}'.  The
great simplification arises since the primitive register type is classed as `always-ready' meaning its {\tt ready} handshake wire for both the {\tt \_read}
and {\tt \_write} methods is implicitly always logical true and needs not be manifest in the RTL.

But in general, every method call has a {\tt ready} net and a rule can only fire if all of the participating methods are ready.  This condition is called
the `{\tt implicit guard}' and the Bluespec compiler creates a conjunction (which in RTL terms is an {\sc AND} gate) of
all of these ready signals and also the explicate guard condition.

The final clause in the conjunction is a schedulling fairness condition generated by the Bluespec compiler's scheduller.  A given method
can appear in a number of rules or several times in a single rule with different arguments.  Where
a non-ephemeral method is invoked with different arguments, only
one can be served at once, since the method corresponds to a physical bus on a physical component that cannot be replicated.  Hence
rules compete with each other and a decision procedure is required.  The standard scheduler is a stateless hardware arbiter that receives requests from each otherwise ready-to-fire rule that contends for a resource and grants just one rule the relevant permission.  See \S\ref{sec:schedullers}.

%----------------------------------------------------------------------------------------

\section{Shortcomings of Guarded Atomic Action Paradigm?}

A purely declarative system cannot update any variables, but, on the other hand, unstructured
multi-threaded imperative code is challenging to understand and reason about.  Hardware is
intrinsically mutable and massively parallel.

Bluespec's rules were motivated by the concept of `guarded atomic
actions' which is a fairly well-known paradigm in computer science.
The idea is that the action body may be expressed in any design style,
including much-loved imperative code, but all of the environment
capture and output side effects occur atomically, as is the case for a
database update or transactional memory system.  In concurrent
systems, explicit control of schedulling order is normally undesirable
since it can lead to excessive serialisation and wastefully reduces
resource utilisation.  Hence, `good system design' tends to be
transactional and robust against the precise transaction order.  But
this cannot always be achieved.  For instance, in the real world, a bank account should
have the same final balance regardless of the order of processing
credits and debits.  But if all the debits happen in the first half of
a year and all of the credits in the second half, many of the debits
may fail to commit owing to an overdraft limit.

In digital hardware, two of the most important components are the register and the RAM.
Both of these are highly sensitive to the precise ordering of read and write transactions.
The three specific behaviours in imperative code that can lead to non-deterministic results are known as RaW, WaR and WaW hazards.
In Bluespec programs, a number of rules are likely to be able to fire at once and indeed the compiler packs some number of these
into a single clock cycle using schedulling rules. The observed result must correspond with
some nominal firing order within the clock cycle (the {\em serialisable} semantic). Where they cannot fit in a single clock cycle, owing to
resource conflicts or various other side conditions (schedulling overrides, timing closure, name alias, complexity ...)
they will either suffer total starvation or fire over
subsequent clock cycles in some order.   The detailed behaviour of the compiled program will
depend on the chosen ordering.  Moreover, important aspects of the overall behaviour may typically be influenced, as in the overdraft example above.

It is my impression that {\bf Bluespec does not really help the system designer manage global ordering artefacts} and that {\bf a Bluespec
  program is fragile because a small change somewhere may influence the firing order elsewhere in the program giving a different observable
  behaviour.}  The difference in behaviour may just be a performance degradation. But it may also be a correctness issue: for instance, the
final result is different under a WaW hazard with alternative resolutions.

The elegance of the guarded atomic action is preserved if registers and RAMs are avoided.  Instead, FIFO and other queue style interfaces
must then be used throughout.  However, designers of real hardware want to use registers and RAMs for efficiency and to express their design ideas.

\subsection{Modified standard semantics}

Bluespec normally requires that the implicit guards of all expressions
hold before a rule can fire.  This can be too strict when non-strict
operators are present.  The logical connectives {\tt ||} {\tt \&\&}
and {\tt ?:} are non-strict in most languages. Should they be non-strict
in Bluespec?  The Bluespec situation is more complex than most high-level languages
owing to the intrinsic guard on a read operation that is not {\tt always-ready}.

So for a rule containing an expression  {\tt (a.x()) ? (b.x()) : 0} should the rule only fire when the
guard for {\tt b.x} holds, or can it fire also when {\tt a.x()}
returns false?  Bluespec solves this by giving the user the choice of
two semantics.  Clearly, using the non-strict variant generally
facilitates less blocking.  But where the transaction is side-effecting
there is a difference as to which side effects occur.  What if the
value of {\tt a.x()} can only be observed at the expense of making
something else un-observable, such as reading different locations from
a register file?\footnote{I say register file here and not RAM because in many FPGA technologies and applications, register files may have combinational read ports whereas RAMs may have synchronous reads and in standard Bluespec\footnote{We have overcome this limitation in a compiler extension to be documented elsewhere.} such RAMs require the whole Put/Get framework.}
A single read port on a register file cannot serve two registers at once, but with the strict semantics, it would potentially be tied up even
when the result was `obviously' not going to be used.  By `obvious' we mean by inspection of a net that was stable early in the clock cycle.

\section{Load Balancing and Schedulling Decisions \label{sec:schedullers}}

In the context of schedulling, the term `fairness' refers to every participant getting at least some service, regardless of how little compared with others.

The fragment shows how users can control the relative precedence of rules using a `{\tt descending\_urgency} annotation.

{\small\begin{verbatim}
(* descending_urgency = "resetCounter, incrementCounter" *)
 
rule incrementCounter;
    action counter <= counter + 1; endaction
endrule

// Next rule resets the counter to 1 when it reaches its limit.
rule resetCounter (counter >= 3);
    action counter <= 1; endaction
endrule

\end{verbatim}}

%       /* The descending_urgency attribute will indicate the scheduling order for the indicated rules. */
This fragment shows two rules that race to update a shared register
and whose ordering is important.  The standard semantics, that allow
only one update to a register per clock cycle, spot the conflict
between the rules and give a higher priority to \verb+resetCounter+
since it has a tighter guard than the increment rule, which is
unguarded.  The {\tt descending\_urgency} annotation has no effect
since the standard scheduller will chose this ordering anyway.  The
behaviour would be for the counter to cycle with a 0, 1, 2, 3, 1, 2,
3, 1 ... sequence.

With our own relaxation of the at most one update per clock cycle
rule, both rules will fire. The resultant
observable pattern depends on the order of composition, which can be
controlled with an {\tt execution\_order} annotation.  If the increment
is placed second in the clock cycle, we see the pattern 0, 1, 2, 3, 2,
3 ... and if placed we see 0, 1, 2, 1, 2.
If one still simply increments and the other now simply decrements, by any amount, the order of composition would not matter.

The concept of {\em fairness}, from formal automaton theory, when applied to Bluespec rules,
would denote that every rule has the opportunity to fire at some point in the future.  The standard Bluespec compilation mode
does not provide fairness.  Instead, the compiler issues a warning where it detects that a rule can never fire and it is
up to the user to modify other rules to be less greedy or to manually instantiate an arbiter.
A Bluespec scheduller implements an automated fairness optimiser.  It schedules all rules in the current compilation
unit but does not countenance fairness between units.

Our implementation partitions the global set of rules into equivalence classes where each rule in a class
has at least one conflicting resource  use with another member of the class.
A conflicting resource pattern is any structural or sharing hazard that requires two different values on the same
physical bus during one clock cycle.
Ephemeral methods are elaborated as many times as needed and do not create sharing conflicts in themselves, only through their resource use.
Also, where a resource is {\tt always\_ready} and has no arguments it does not present a sharing conflict.

Schedulling decisions can then be made for each class in isolation.  Within a class, we consider the composite
guards of the rules in conjunctive normal form. These may have items of support (clauses) in common with the same or complementary polarity.  Any pair with clauses of complementary polarity are clearly mutually exclusive and can be schedulled separately.

Where priority cannot be granted based on exclusion, a static priority is needed since automatic instantiation of
stateful arbiters is not part of the standard semantic.
Any pair with common clauses may find that one strictly implies the other.  For instance {\tt a.b.c } implies {\tt a.b} (where dot denotes conjunction).  Hence priority should be allocated to those with the longer conjunction over those that can manifestly fire more frequently. But this may be overriden with the {\tt descending\_urgency} or {\tt fire\_when\_enabled} user annotations.
In the remaining circumstances, arbitrary yet consistent resolution is required and we choose to do this
based on a lexicographical sort of the file name and line number were the rule is defined.

A rule whose composite guard is manifestly false will never fire.  This is reported as a compile-time warning of high severity to the user.  A rule can still suffer total starvation in practice even if this is not spotted by the compiler: for instance it might simply depend on an external input pattern that never occurs in the final circuit's surrounding context. Or it might depend on the system being in a state that is actually unreachable since no pattern of inputs cause that sate to be entered: this would requires a liveness modelcheck to report.

\subsection{Stateful and other Alternative Schedullers}

The arbitration described in the previous section is roughly equivalent to what the commercial Bluespec compiler implements (as far as we know).
In particular, it does not generate any additional state bits that would be needed by, for example, a round-robin arbitration system.  The suggested design approach seems to be for the user to look at starvation warnings generated by a compilation run, and if not happy, to manually instantiate an arbiter.  This is very easy to do with the Bluespec syntax: the arbiter needs to have sufficient methods for each of the contesting rules to include one method call in its workings, such as in its explicit guard.  The method does not need any argument or result: just calling it is sufficient owing to the  Bluespec implicit guard semantics, owing to it having an implicit guard.

However, an alternative approach is for the Bluespec compiler to implement its own multi-cycle scheduller automatically.  Surely this is more user-friendly?  The results of our experiments should be in a to-be-published paper based on this preprint.

\section{Conclusions and Further Work \label{sec:conclusion}.}

Bluespec is much lower level than C-to-gates HLS tools such as LegUp\cite{LEGUP13}
and Kiwi\cite{Kiwi08}.  Bluespec's constraint that all rules must complete in a
single cycle rules out simple access to multi-cycle primitive
components.  Important multi-cycle components are synchronous static
RAMs (BRAM in FPGA) and the multipliers (DSP blocks in FPGA).  For access to
these resources, the Bluespec Put/Get interface must be used, where the
arguments are supplied by one rule firing and the results collected by
a later firing of the same or another rule.  HLS tools generate
multi-cycle schedules based on the dataflow within a basic block (or
catenation of multiple blocks from loop unwinding).

Moreover, multi-cycle components, such as our multiplier, must normally be explicitly named in the Bluespec source code with
the binding of an operation to a specific instance being manual.  HLS tools perform automatic load balancing and
bind to operators that have wiring affinity, permuting operands to commutative ALUs where possible to reduce multiplexor count.
A multi-cycle scheduler that instantiates arbiters and sequencers and that can address hard real-time performance targets
seems to be an obvious next step.

The standard compiler does not render RTL with a parameterised top-level.  For features such as
databus width or constant initialisation (eg.~component serial number), this is not a hard feature to add
so we will do it.  It would able be possible to emit RTL containing a parameterised {\tt generate} statement that
enables as many instances of a method to elaborated as desired, thereby solving the loss of ephemeral methods
over compilation unit boundaries.  This could be useful for switch and bus infrastructure.

% Modular scheduling of guarded atomic actions Daniel L. Rosenband, Arvind 

Download open source tarball \htmladdnormallink{\tiny https://www.cl.cam.ac.uk/~djg11/wwwhpr/toy-bluespec-compiler.html}{https://www.cl.cam.ac.uk/~djg11/wwwhpr/toy-bluespec-compiler.html} or email for git access to live repo.
\bibliographystyle{IEEEtran} \bibliography{opensrc-bluespec-greaves} 

% Like Chisel, it has good support for {\bf valid-tagged data} in registers and busses.  Hence compiler
% optimisations that ignore dead data are potentially possible.

This reseach note document was first circulated as a pre-print in 2013 and has been re-issued early 2019 with small edits
after short discussions with Jonathan Woodruff and Jamey Hicks.
\end{document}